\documentclass[journal=ancac3,manuscript=article]{achemso}

\usepackage[version=3]{mhchem}

\author{Sung Hwan Kim}
\affiliation{Center for Artificial Low Dimensional Electronic Systems, Institute for Basic Science (IBS), Pohang 37673, Republic of Korea}
\alsoaffiliation{Department of Physics, Pohang University of Science and Technology (POSTECH), Pohang 37673, Republic of Korea}

\author{Kyung-Hwan Jin}
\affiliation{Department of Materials Science and Engineering, University of Utah, Salt Lake City, Utah 84112, USA}

\author{Byung Woo Kho}
\affiliation{Department of Physics, Pohang University of Science and Technology (POSTECH), Pohang 37673, Republic of Korea}

\author{Byeong-Gyu~Park}
\affiliation{Pohang Accelerator Laboratory, Pohang University of Science and Technology (POSTECH), Pohang 37673, Republic of Korea}

\author{Feng~Liu}
\affiliation{Department of Materials Science and Engineering, University of Utah, Salt Lake City, Utah 84112, USA}
\alsoaffiliation{Collaborative Innovation Center of Quantum Matter, Beijing 100084, China}

\author{Jun Sung Kim}
\affiliation{Department of Physics, Pohang University of Science and Technology (POSTECH), Pohang 37673, Republic of Korea}

\author{Han~Woong Yeom}
\email{yeom@postech.ac.kr}
\affiliation{Center for Artificial Low Dimensional Electronic Systems, Institute for Basic Science (IBS), Pohang 37673, Republic of Korea}
\alsoaffiliation{Department of Physics, Pohang University of Science and Technology (POSTECH), Pohang 37673, Republic of Korea}

\title{Atomically Abrupt Topological \textit{p-n} Junction}

\keywords{topological insulator, topological \textit{p-n} junction, angle-resolved photoemission spectroscopy, scanning tunneling microscopy/spectroscopy, ultrathin Sb film}

\begin{document}

\begin{abstract}

\begin{figure}[h]
\includegraphics[width=0.5\linewidth]{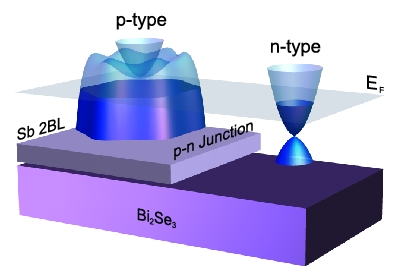}
\end{figure}
\textbf{ ABSTRACT:
Topological insulators (TI's) are a new class of quantum matter with extraordinary surface electronic states, which bear great potential for spintronics and error-tolerant quantum computing. In order to put a TI into any practical use, these materials need to be fabricated into devices whose basic units are often \textit{p-n} junctions. Unique electronic properties of a 'topological' \textit{p-n} junction were proposed theoretically such as the junction electronic state and the spin rectification. However, the fabrication of a lateral topological \textit{p-n} junction has been challenging because of materials, process, and fundamental reasons. Here, we demonstrate an innovative approach to realize a \textit{p-n} junction of topological surface states (TSS's) of a three-dimensional (3D) topological insulator (TI) with an atomically abrupt interface. When a ultrathin Sb film is grown on a 3D TI of Bi$_2$Se$_3$ with a typical $n$-type TSS, the surface develops a strongly $p$-type TSS through the substantial hybridization between the 2D Sb film and the Bi$_2$Se$_3$ surface. Thus, the Bi$_2$Se$_3$ surface covered partially with Sb films bifurcates into areas of $n$- and $p$-type TSS's as separated by atomic step edges with a lateral electronic junction of as short as 2~nm. This approach opens a different avenue toward various electronic and spintronic devices based on well defined topological \textit{p-n} junctions with the scalability down to atomic dimensions.
}
\end{abstract}

Surface states of topological insulators\cite{rmp.82.3045, rmp.83.1057}, called topological surface states (TSS's), are robustly protected by the bulk topological nature and form necessarily a Dirac band with their spins locked helically with momentum\cite{n.452.970, s.325.178, rmp.82.3045, rmp.83.1057, prl.105.146801, np.5.438}. These unique properties find obvious merits in spintronic applications and can yield a Majorana Fermion in proximity with superconductivity\cite{s.301.1348,prl.92.126603}. However, there has been a huge barrier in making devices based on TSS's. The challenge is closely related to the notorious issue of controlling impurities or dopants in a TI crystal. While quite a few works tried to control the chemical potential of a TSS by impurity doping\cite{n.460.1101, s.325.178, prl.104.016401, np.7.32, np.8.616, s.329.659, s.339.1582}, the deterioration of the surface channel and the inclusion of bulk channels were inevitable in many cases. Especially, the tunability of the chemical potential was often not enough to make a good $p$-type TSS. Fabricating a well defined topological \textit{p-n} junction is even more challenging\cite{am.25.889, acsnano.9.10916, am.28.2183, nc.7.13763}, which represents one of the most important technological issues in staging applications of TI's. Nevertheless, a topological \textit{p-n} junction, defined as an electronic junction of a \textit{p-} and a \textit{n-}type TSS, features unique properties, which are not shared by conventional \textit{p-n} junctions of semiconductors but promise attractive new applications\cite{s.301.1348,prl.92.126603}. At a topological \textit{p-n} junction, the electron scattering and transport are largely governed by the spin polarization of TSS's involved. This property provides the spin rectification effect and a few other spintronic applications\cite{prl.115.096802,prl.114.176801}. Moreover, under a magnetic field an 1D electronic channel develops along the junction.\cite{prb.85.235131}

 Due to layered structures of most of 3D TI crystals, the stack of $n$- and $p$-type TI crystals is feasible only in the vertical direction \cite{nc.6.8816}. The laterally graded bulk doping was only applied with a partial success to create a junction between a $n$- and a marginally $p$-type TI. However, the \textit{p-n} junction of its surface channel was not clearly established and the junction length is limited to 40 nm\cite{am.28.2183}. A very recent work used the surface doping by an organic molecular film to create successfully a topological \textit{p-n} junction, which would have a similar length scale \cite{nc.7.13763}. In general, the poor dielectric screening in 2D or layered materials can impose a fundamental limit to the junction length \cite{nl.16.5032}. 

\begin{figure}[!pt]
\includegraphics[width=0.5\linewidth]{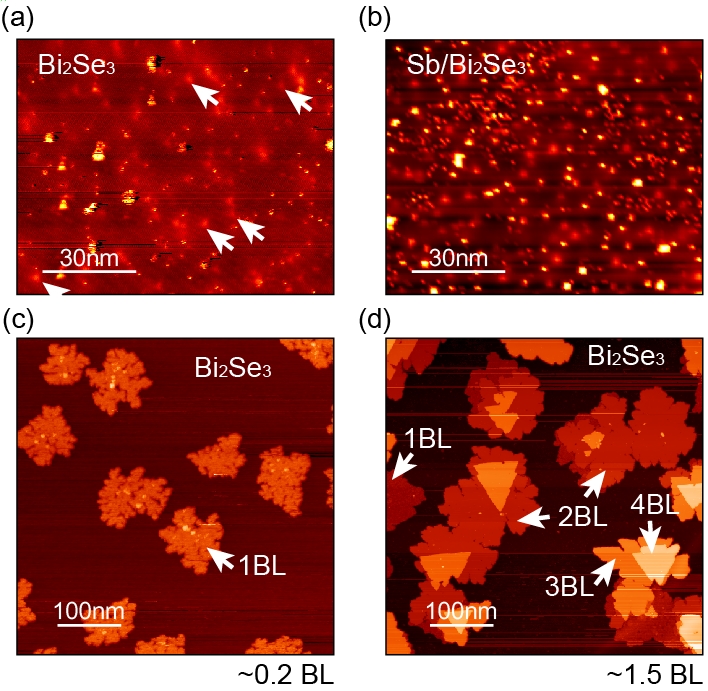}
\caption{STM images for the initial growth of Sb film. The STM topography of (a) clean and (b, c, d) Sb-deposited Bi$_2$Se$_3$ surfaces. The characteristic defects are resolved on the bare Bi$_2$Se$_3$ surface [indicated by arrow in (a)]. (b) At very low coverage, Sb atoms and small cluster less than 1 nm in size are scattered on the surface. (c) Near 0.2 bilayer (BL) coverage, Sb islands of 1 BL height are formed but their surfaces are rough. (d) At a higher coverage about 1.5 BL, well-ordered Sb islands appears, whose size is a few hundred nm and height is 2$\sim$4 BL's.}
\end{figure}

On the other hand, there is a radically different way discovered recently to engineer, in principle, a TSS with impurity and defect issues. While the existence of a TSS is guaranteed and its electronic states are immune to external pertubations, the dispersion of a TSS depends strongly on chemical and geometrical structures of the surface. Various theoretical works predicted that the dispersion and the number of edge or surface state bands for a 2D or 3D TI change drastically upon choosing different atoms to terminate its surface \cite{prl.110.016403, prb.90.165412, nl.14.2879, pccp.18.8637, prb.93.075308, sr.3.3060, nl.13.6064, prl.115.136801, prb.90.155414}. Indeed, recent ARPES and scanning tunneling spectroscopy (STS) works showed that TSS's with totally different dispersions and spin textures are formed by the Bi film termination of Bi-based ternary chalcogenides \cite{prl.107.166801,prl.109.016801, pnas.110.2758, nc.4.1384, prb.89.155436, prb.93.075308, nm.14.1020, jpcm.28.255501, acsnano.10.3859}. This phenomenon was called the TSS transformation \cite{prb.89.155436, prb.93.075308} or the topological proximity effect \cite{nc.6.6547}, where the hybridization of electronic states of the terminating film and the TSS plays a crucial role.  

We take advantage of the TSS transformation to fabricate an atomically abrupt topological \textit{p-n} junction. Well-ordered double-layer films of Sb are grown on top of a 3D TI Bi$_2$Se$_3$, which yield a strongly \textit{p}-type TSS through the hybridization of 2D Sb bands with the surface state of Bi$_2$Se$_3$. The $p$-type TSS formed is identified by angle-resolved photoemission spectroscopy (ARPES) and scanning tunneling spectroscopy (STS) as predicted by \textit{ab initio} calculations. When the Sb film partially covers the Bi$_2$Se$_3$ surface with a typical \textit{n}-type TSS, a well defined topological \textit{p-n} junctions form naturally at the atomically abrupt edges of Sb islands. This work demonstrates the unambiguous realization of a \textit{p-n} junction of a TSS with an extremely short junction length of only 2 nm.

\begin{figure}[!pt]
\includegraphics[width=0.5\linewidth]{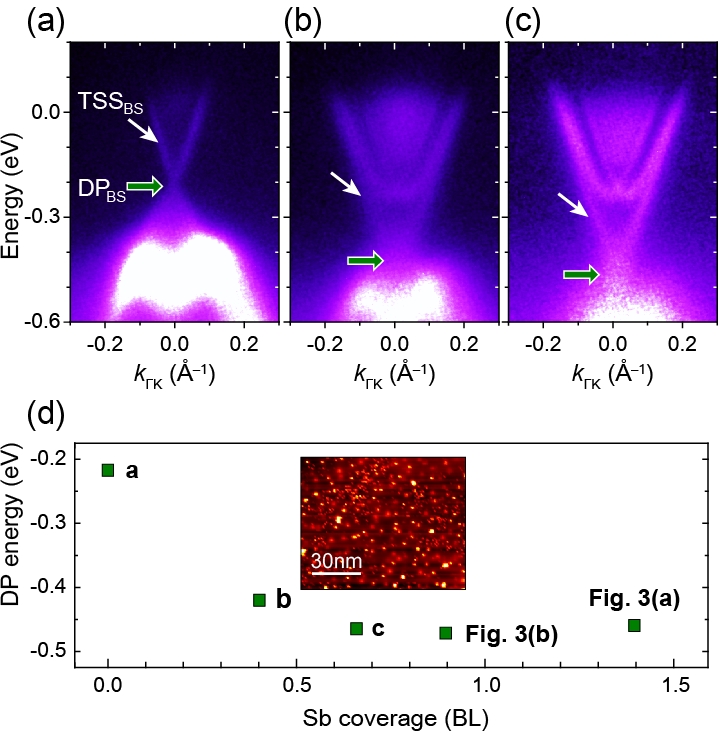}
\caption{Topological surface states of Bi$_2$Se$_3$ without and with Sb submonolayer films. ARPES measurements of surface electronic band dispersions of Bi$_2$Se$_3$ along the $\overline{\Gamma}$--$\overline{\textrm{K}}$ Brillouin zone (a) without Sb and (b) and (c) with Sb of a nominal thickness of 0.4 and 0.67 BL, respectively. Topological surface states of Bi$_2$Se$_3$ (TSS$_\textrm{BS}$) and its Dirac points (DP$_{\textrm{BS}}$) are indicated by white and green arrows, respectively. The change of Dirac points energy with respect to the coverage of Sb is summarized in (d).}
\end{figure}

\section{RESULTS AND DISCUSSION}

Bi$_2$Se$_3$ is taken as a well established 3D TI to be fabricated into a topological \textit{p-n} junction (Figure~1a). High resolution ARPES measurements (Figure~2a) show vividly the TSS of a clean Bi$_2$Se$_3$ crystal cleaved in ultra-high vacuum. Its Dirac point (DP$_{\textrm{BS}}$) is located near $-$0.2 eV indicating an electron-doped surface. The Dirac point energy is further confirmed by STS measurements as shown in Figure~S1 of Supporting information. This doping effect has been attributed to the characteristic defects, which were resolved clearly in previous and present scanning tunneling microscopy (STM) topographies (Figure~1a)\cite{prb.69.085313, prb.79.195208, prl.108.206402}. After the deposition of Sb below 0.7 BL (nominal thickness) at room temperature, the whole bands of Bi$_2$Se$_3$ are shifted rigidly to a lower energy (Figures~2b and 2c). It is obvious that deposited Sb atoms donate extra electrons to the surface. They also generate a potential gradient into the subsurface region, which causes quantum-well type 2D electron gas in the conduction band (Figure~2c). Several other metals such as Fe, In, and Cu have been reported to cause the same effect.\cite{np.7.32, nc.3.1159, prb.91.121110}. However, such doping effect saturates with the Dirac point  of TSS at $-$0.47 eV (Figures~2d and 3b) for the Sb deposition beyond a nominal thickness of 0.7 bilayer (BL). This is due to the onset of the aggregation of Sb adatoms into compact 2D islands as revealed by STM images at this coverage (Figures~1c, 1d, and 3a). In contrast, at a lower coverage, they spread over the surface as adatoms or small clusters of size less than 1 nm (Figure~1b). Note that the lattice mismatch between Bi$_2$Se$_3$ and Sb(111) is so small as 0.36~\% to ensure a good epitaxial relationship\cite{sr.6.33193, prb.93.075308}. At a higher nominal coverage of about 0.9 BL, roughly half of the surface is covered by Sb islands of a width of a few hundred nanometers and the majority (60 $\sim$ 70~\%) of them have a height of 2 BL ($\sim$1 nm) with flat tops.

At this high coverage regime (see Figure~2d), ARPES measurements show a set of extra bands (TSS$_{\textrm{Sb}}$ and B$_1$ bands in Figure~3b), which can straightforwardly be attributed to Sb islands. Most importantly, a metallic surface state (TSS$_{\textrm{Sb}}$) appears to disperse linearly towards the center of Brillouin zone. The other extra band B$_1$ is rather faint but discernible to disperse adjacent to TSS$_{\textrm{Sb}}$ and TSS$_\textrm{BS}$ with its parabolic band top around $-$0.05 eV. This mixed band structure with two metallic surface states dispersing differently is further shown by constant energy contours at several energies in Figure~3c. Its detailed dispersion can be found more clearly in the momentum and energy distribution curves extracted from Figure~2b (Figure~3d and Supporting Information Figure~S2).

\begin{figure}[!pt]
\includegraphics[width=1.0\linewidth]{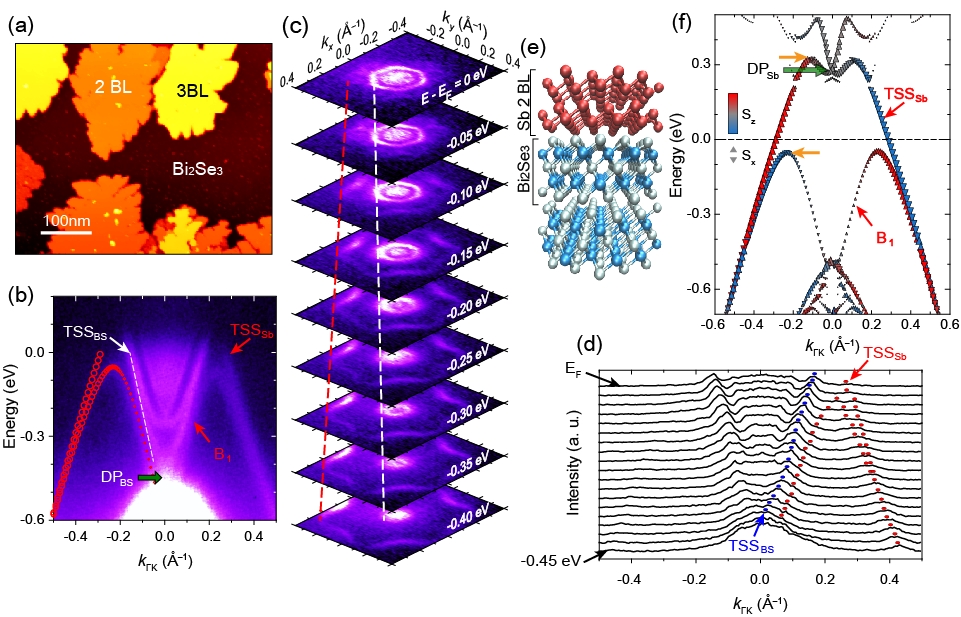}
\caption{Development of a topological surface state for compact Sb islands on Bi$_2$Se$_3$. (a) STM image for  patially covered Sb on Bi$_2$Se$_3$. (b) ARPES measurements along $\overline{\Gamma}$--$\overline{\textrm{K}}$ of Bi$_2$Se$_3$ covered roughly half by Sb films, whose STM image is shown in (a). (c) Constant energy contours of the ARPES intensity in 2D Brillouin zone at the energies specified. A metallic state (TSS$_\textrm{Sb}$ and the red dashed line) is noticeable together with the TSS of Bi$_2$Se$_3$ (TSS$_\textrm{BS}$ and while dashed lines), which are also well characterized in momentum distribution curves (MDCs) in (d) (indicated by red and blue dots, respectively). (f) Calculated band structure of a 2 BL Sb film on Bi$_2$Se$_3$ with its spin textures, where the structure model shown in (e) is used.}
\end{figure}

The origin of the extra bands can be unambiguously understood from \textit{ab initio} density functional theory calculations\cite{njp.12.043048, prb.81.115407, prb.93.075308}. The calculations are performed for the structure with a Sb 2 BL film on top of a 6 quintuple layer Bi$_2$Se$_3$ substrate (Figure~3e). The resulting band structure on the Sb layer (Figures~3f, 4b, and 4c) has a metallic Dirac band of TSS$_{\textrm{Sb}}$ and a fully occupied B$_1$ band. The TSS$_{\textrm{Sb}}$ has its Dirac point at $+$0.27 eV above E$_\textrm{F}$ and is thus strongly \textit{p}-type. As shown in Figures~3f, 5e, and 5f, these two bands are fully spin polarized. The spin texture and the dispersion of B$_1$ indicate that it is a Rashba-type band. As compared in Figure~3b, the calculated band dispersions (open circles) match perfectly with those of TSS$_{\textrm{Sb}}$ and B$_1$ measured by ARPES. This band structure is essentially consistent with those of single Sb and Bi BL films on various TI substrates, which were analyzed fully in the previous works\cite{prb.90.235401, prb.93.075308}. That is, the topmost valence band of Sb overlaps and hybridizes with TSS$_\textrm{BS}$ to split into these two bands. The original surface state of TSS$_\textrm{BS}$ is hybridized into B$_1$ band and looses its topological character (Figure~4). Instead, the metallic band TSS$_{\textrm{Sb}}$, localized mostly within the Sb film, takes the topological role for the Sb-terminated Bi$_2$Se$_3$ surface. 
While the surface measured in ARPES contains a minor area of 3 or 4 BL films of Sb, their contributions in the ARPES signal are not noticed even though the 3B film has a distinct band dispersion (Supporting information Figure~S3). 

The heterostructure of Sb and Bi$_2$Se$_3$ can be interpreted as a combination two different 3D TI's where two different surface states appear; top and bottom TSS's from the different TI's, respectively (Figures~4d and 4e). The topological nature of a thin Sb film is lost due to the coupling between top and bottom surface states but recovers when its bottom surface state hybridizes with the TSS of Bi$_2$Se$_3$ to break the coupling between TSS's of the film. As a result, the emerging TSS comes solely from the top Sb bilayer of the film and its strongly $p$-type character is due to the charge transfer from Sb to Bi$_2$Se$_3$ (Figure 4f). This idea is solidly confirmed in by solving our model Hamiltonian and through detailed band structure calculations (Supporting information). Therefore, we can unambiguously conclude that the Bi$_2$Se$_3$ surface covered partially with Sb has simultaneously \textit{n}- and \textit{p}-type TSS's in the bare surface and the Sb-covered surface, respectively (Figure 4g). It is straightforward to expect a \textit{strong} \textit{p-n} junction at the boundary between ultrathin Sb films and the bare Bi$_2$Se$_3$ surface since the \textit{n}- and \textit{p}-type Dirac points are separated by a huge energy scale at $-$0.45 and $+$0.27~eV, respectively (Figures~3b and 3f). 

\begin{figure}[!pt]
\includegraphics*[width=1.0\linewidth]{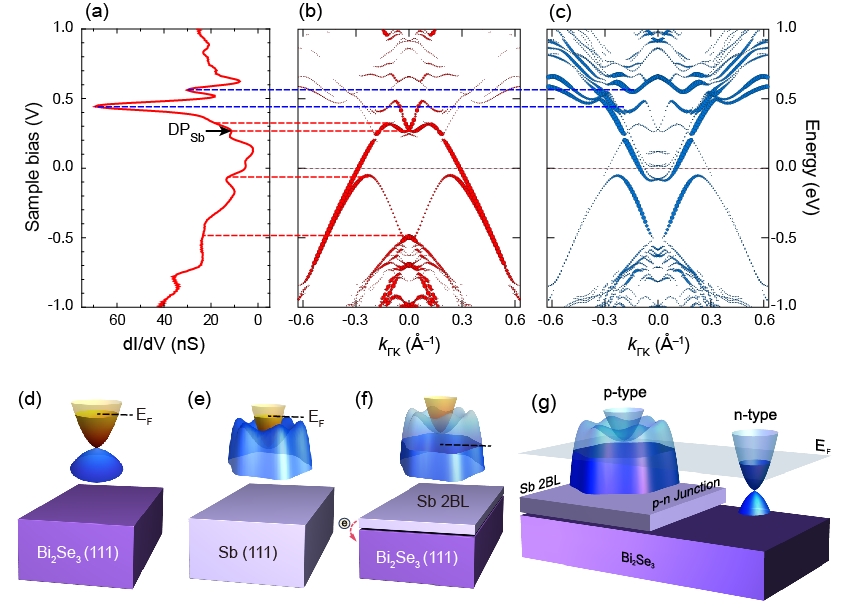}
\caption{Origins of the Sb film's STS spectral features and schematics of surface states for Bi$_2$Se$_3$ and Sb 2 Sb BL/Bi$_2$Se$_3$. (a) The STS spectrum measured inside of 2 BL Sb film. The calculated band structure originated from (b) the 2 BL Sb film and (c) the top quintuple layer of the Bi$_2$Se$_3$ substrate, which is taken from the \textit{ab initio} calculation for the structure of Figure~3e. The STS spectrum (a) fits well with the calculated band structures. All the spectral features come from the hybridization between the Sb film and the substrate. While the two peaks near $+$0.5 eV, indicated by blue dashed lines, mainly come from the substrate, the peaks near $+$0.3 and $-$0.5 eV largely from Sb. Schematics of surface states for (d) Bi$_2$Se$_3$(111), (e) Sb(111) and (f) 2 BL Sb/Bi$_2$Se$_3$(111). The TSS's for Bi$_2$Se$_3$ and Sb are typically $n$-type. In Sb/Bi$_2$Se$_3$(111), the original surface state of Bi$_2$Se$_3$ hybridizes with the bottom surface state of Sb film and is buried into bulk states. Thus, the TSS emerges solely from the top Sb bilayer. The energy and shape of the Dirac cone are slight different from the pristine Sb(111) TSS but its helical nature is well preserved. (g) Schematic diagram for the topological \textit{p--n} junction formed on Bi$_2$Se$_3$ with a partially grown Sb film.}
\end{figure}

The lateral \textit{p-n} junctions at edges of Sb islands can be directly confirmed and visualized in atomic scale by STS measurements (Figure~5). The $dI/dV$ curves of the tunneling current $I(V)$ are obtained across an edge of a Sb 2 BL film (Figure~5a, black arrow), which shows local density of states near the edge. The representative  $dI/dV$ curves of Bi$_2$Se$_3$ (Figure~5d, violet solid line) is consistent with previous reports\cite{prl.105.076801, np.10.294}, and its Dirac point (DP$_{\textrm{BS}}$) is located near $-$0.22 eV (DP$_{\textrm{BS}}$, the lowest LDOS point of the $dI/dV$ curve). The location of DP$_\textrm{BS}$ is different from the ARPES measurement (Figure~3b) by about 0.2~eV as caused by the lateral band bending near the Sb island. The DP$_\textrm{BS}$ away from the Sb islands is confirmed to agree with the ARPES result (See Supporting Information Figure~S1). On the other hand, the $dI/dV$ curves on the Sb island (Figure~5d, red solid line) reflect its drastically different band structure. The most prominent features at $+$0.45 and $+$0.55 eV are assigned through comparison with the calculation to the hybridized states in the conduction bands (indicated by blue arrows in Figures~4a, 4b, 4c, and 5d). The band edges of TSS$_{\textrm{Sb}}$ appear with smaller intensities at about 0.3 and $-$0.1 eV (as indicated by orange arrows) (Figures~4a, 4b, 4c, and 5d). The comparison with the calculation further indicates the position of the formed Dirac point (DP$_\textrm{Sb}$) near $+$0.25~eV (Figures~~4a, 4b, 4c, and 5d). 

The lateral \textit{p-n} junction is sustainable for a smaller island until the edge effect destroys the 2D band structure of the Sb film and its interface. The edge effect is expected to be substantial for an island size smaller than 10 nm as estimated from the lateral extent of the edge state of the Sb film. Within our calculation, the \textit{p-n} junction or the  \textit{p-}type TSS is also preserved for a thicker film up to a thickness of 5 BL's while the electronic structure and the junction property become more complicated for a thicker film.   

\begin{figure}[!pt]
\includegraphics*[width=1.0\linewidth]{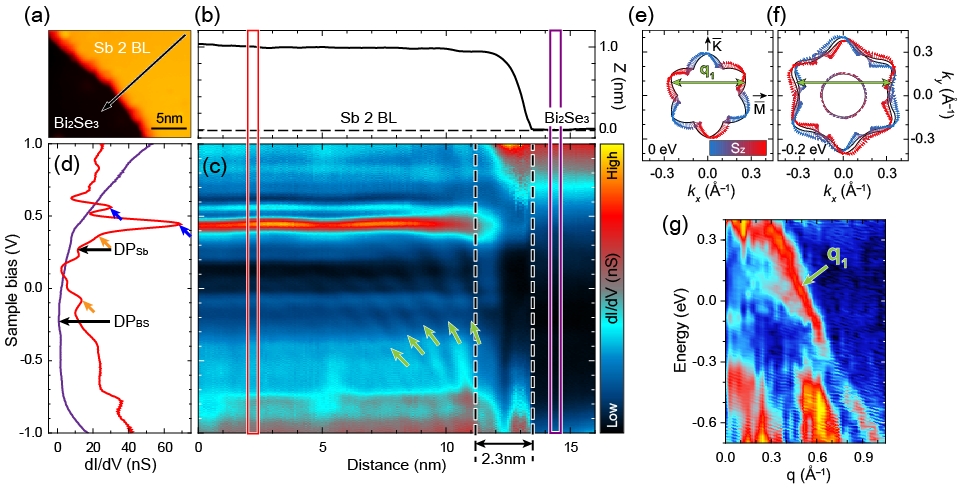}
\caption{Local density of states for the 2 BL Sb film and its edge measured by STM. (a) STM topography of an edge of a 2 BL Sb island and (b) its height profile obtained together with the $dI/dV$ measurement. (c) 2D plot of the $dI/dV$ line scan along the arrow in (a) crossing the edge. Quasi-particle interference (QPI) is observed from the edge into the Sb film as indicated by green arrows. The electronic transition region across the edge is as narrow as 2.3 nm. (d) Typical $dI/dV$ curves taken from the inner side of Sb 2 BL's film and the bare Bi$_2$Se$_3$ surface as sampled in the red and violet rectangles, respectively. The spin textures of the surface state bands of the Sb/Bi$_2$Se$_3$ system at (e) E$_\textrm{F}$ and (f) $-$0.2 eV. The possible scattering vectors (\textbf{\textit{q$_1$}}) connecting the same spin momenta are indicated. (g) Energy-resolved Fourier transform of (c) shows the QPI scattering wave vector \textbf{\textit{q}}$_1$.}
\end{figure}

The key features of the electronic structure of the Sb-covered surface are the formed metallic band TSS$_{\textrm{Sb}}$ and its Dirac point. These features can further be confirmed by detailed $dI/dV$ measurements, which exhibit the quasi-particle interference (QPI) within the Sb island. The QPI due to back scatterings at the edge of the island can be noticed as a wavelike pattern in the $dI/dV$ map (green arrows in Figure~5c). The Fourier transform of the QPI pattern reveals a well defined scattering wave vector \textbf{\textit{q$_1$}} across E$_\textrm{F}$ (Figures~5e, 5f, and 5g). This evidences a single metallic band as predicted on the Sb film by the calculation. The linear dispersion and the \textit{p}-type character, \textbf{\textit{q$_1$}} increasing for a lower energy, are apparent. This result is fully consistent with the ARPES result and the calculation (Figures~3b and 3f). The Dirac point, where \textbf{\textit{q$_1$}} approaching to zero, is estimated to be around roughly 0.3 eV in consistency with the calculation. Ideally, a TSS does not allow the backscattering requested for QPI due to its spin-momentum locking property, but a TSS in a real TI crystal does because of the spin reorientation by the warping of the TSS band \cite{prb.90.235401}. The warping effect on the spin texture is very prominent in our calculation for TSS$_{\textrm{Sb}}$ (Figures~5e and 5f); there arise strong out-of-plane spin components (S$_\textrm{z}$), which flip at $\overline{\textrm{M}}$ direction with in-plane spins kept helical (Figures~5e and 5f). This distorted spin texture allows one scattering vector \textbf{\textit{q$_1$}} connecting spins of the same direction (Figures~5e and 5f, green arrows) as observed experimentally. 

\section{CONCLUSION}

The ARPES and STS experiments show unambiguously that different types of TSS's exist in Sb-covered and Sb-free areas, a topological \textit{p-n} junction is formed between these two areas.  What is very important is that the junction is defined by only a single atomic step, which is the most abrupt junction possible in a crystal. This corresponds to an ideal case of an abrupt step function doping profile. The \textit{electronic} junction would be wider than the single atom step and its width can be experimentally measured from STS maps (Figure~5c). Nevertheless, it is still as short as 2.3~nm (black arrow between dashed lines in Figure~5c). Since this junction is not based on accumulated charges as in the cases of doping or gating, the dielectric screening, which limits the junction length significantly in 2D or layered systems\cite{nl.16.5032}, would not be important in the present case. That is, the distinct feature of the present junction is based on its fundamentally different mechanism, the transformability of a TSS by surface termination. Note also that the electron densities of the \textit{n}- and \textit{p}-type regions are occasionally quite similar as estimated from similar Fermi vectors of the corresponding TSS's. This is a desirable characteristics for a good \textit{p-n} junction. Moreover, it is not difficult to find a \textit{n-p-n} or \textit{p-n-p} junction, either, when crossing a single Sb island or a two neighboring islands. In addition, the distinct spin texture of the present TSS (Figure~5e) might be exploited further for spin transports\cite{s.301.1348,prl.92.126603}. Therefore, a distinct approach to create atomically abrupt topological \textit{p-n} junction has been demonstrated here, which would definitely accelerate the development of various electronic/spintronic devices of TSS's, scalable down to atomic dimensions, and of fundamental researches on exotic electronic properties of topological junctis.

\section*{Methods}

\textbf{Sample Growth.} The single crystals of Bi$_2$Se$_3$, used as substrates, were grown using self-flux method and cleaved \textit{in vacuo}. The Sb films were grown on the Bi$_2$Se$_3$ surface by a thermal effusion cell at room temperature\cite{prb.89.155436, prb.90.235401, sr.6.33193}.

\textbf{STM/STS Experiments.} We performed STM/STS experiments using a commercial low-temperature STM (Unisoku, Japan) in ultrahigh vacuum better than 5$\times$10$^{11}$ Torr at $\sim$78 K. STM topographic data were obtained using the constant current mode. The STS spectra ($dI/dV$ curves) were obtained using the lock-in technique with a bias-voltage modulation of 1 kHz at 10--30 mV$_rms$ and a tunneling current of 500--800 pA.

\textbf{ARPES Measurements.}
The ARPES measurements were performed on \textit{in situ} clean and Sb deposited Bi$_2$Se$_3$ single crystal surfaces with a high performance hemispherical electron analyzer (VG-SCIENTA R4000) at the 4A1 ARPES beamline in Pohang Accelerator Laboratory. The samples were kept near 50 K for measurements. The ARPES data were taken at the photon energy of 24 eV.

\textbf{\textit{Ab initio} Calculations.} 
The \textit{ab initio} calculations were carried out in the framework of generalized gradient approximations with Perdew-Burke-Ernzerhof functional using the plane wave basis Vienna ab inito simulation package (VASP) code \cite{prb.54.11169, prl.77.3865}. All the calculations were carried out with the kinetic energy cutoff of 400 eV on the 11$\times$11$\times$1 Monkhorst-Pack k-point mesh. A vacuum layer of 20\AA-thick was used to ensure decoupling between neighboring slabs. The Sb slabs were fully optimized until the Helmann-Feynman forces are less than 0.01 eV/\AA. Six quintuple layers were used to simulate the Bi$_2$Se$_3$ substrate. The spin-orbit coupling is included in the self-consistent electronic structure calculation.

\section{ASSOCIATED CONTENT}
\subsection{Supporting Information}
The Supporting Information is available free of charge on the ACS Publications website at http://pubs.acs.org.

STS measurements on a Bi$_2$Se$_3$ surface; high resolution ARPES measurement of Sb/Bi$_2$Se$_3$ and its energy distribution curves; comparison of the band structures for 2  and 3 BL Sb/Bi$_2$Se$_3$; theoretical support of the mechanism for the formation of the \textit{p-}type topological surface state by Sb layers (PDF).

\subsection{AUTHOR INFORMATION}
\subsubsection{Corresponding Author}
*E-mail: yeom@postech.ac.kr

\section{Acknowledgements}
S. W. Jung, W. J. Shin and K. S. Kim helped the ARPES measurements.
This work was supported by Institute for Basic Science (IBS) through the Center for Artificial Low Dimensional Electronic Systems (Grant No. IBS-R014-D1), and by the National Research Foundation (NRF) through the Center for Topological Matter (Grant No. 2011-0030785), and the Max Planck POSTECH/KOREA Research Initiative (Grant No. 2011-0031558) programs. K.-H.J. and F.L. acknowledge financial support from DOE-BES (No. DE-FG02-04ER46148) and thank Supercomputing Center at USTC, NERSC and CHPC at University of Utah for providing the computing resources.

\providecommand*\mcitethebibliography{\thebibliography}
\csname @ifundefined\endcsname{endmcitethebibliography}
  {\let\endmcitethebibliography\endthebibliography}{}

\end{document}